\documentclass[twocolumn,showpacs]{revtex4}
\pdfoutput=1
\usepackage{graphicx}
\usepackage{dcolumn}
\usepackage{longtable}
\usepackage{amsmath}
\usepackage{amssymb}

\begin{document}

\title
{Regularity and chaos in $0^+$ states of the interacting boson model using quantum measures}

\author
{S. Karampagia$^1$, Dennis Bonatsos$^1$, and R. F. Casten$^{2}$}

\affiliation
{$^1$ Institute of Nuclear and Particle Physics, National Centre for Scientific Research 
``Demokritos'', GR-15310 Aghia Paraskevi, Attiki, Greece}

\affiliation
{$^2$ Wright Nuclear Structure Laboratory, Yale University, New Haven, CT 06520, USA}

\begin{abstract}

{\bf Background}: Statistical measures of chaos have long been used in the study of 
                  chaotic dynamics in the framework of the interacting boson model. 
                  The use of large number of bosons renders additional studies of 
                  chaos possible, that can provide a direct comparison with similar 
                  classical studies of chaos.

{\bf Purpose}: We intend to provide complete quantum chaotic dynamics at zero angular 
               momentum in the vicinity of the arc of regularity and link the results 
               of the study of chaos using statistical measures with those of the 
               study of chaos using classical measures.

{\bf Method}: Statistical measures of chaos are applied on the spectrum and the transition 
              intensities of $0^+$ states in the framework of the interacting boson model.

{\bf Results}: The energy dependence of chaos is provided for the first time using 
               statistical measures of chaos. The position of the arc of regularity was 
               also found to be stable in the limit of large boson numbers. 

{\bf Conclusions}: The results of the study of chaos using statistical measures are 
                   consistent with previous studies using classical measures of chaos, as 
                   well as with studies using statistical measures of chaos, but for small 
                   number of bosons and states with angular momentum greater than 2.

\end{abstract}


\maketitle
    
\section{Introduction} %1

The Interacting Boson Model (IBM) \cite{IA}, apart from being successful in describing the 
low-lying levels and electromagnetic transition intensities of even-even heavy nuclei, 
has also been used in studying transitions \cite{JPG,RMP} between the different dynamical symmetries of 
the model, by changing its parameters. 
The model is known to possess 3 dynamical symmetries, namely U(5), O(6) and SU(3). 
When a system possesses a certain dynamical symmetry, it is completely integrable. Away 
from these dynamical symmetries, one would expect chaos. However, this is not 
always the case. A situation of great interest is the notion of quasidynamical symmetry (QDS), 
i.e., the approximate persistence of a symmetry, in spite 
of strong symmetry breaking interactions \cite{Rowe1225,Rowe2325,Rowe745,Rowe756,Rowe759}. 
Symmetry breaking can also be seen as the transition of a system from regular dynamics 
(exhibited by the presence of a dynamical symmetry) to chaos \cite{Gutzwiller,Haake}. 

The interplay between regular and chaotic behavior in the context of the Interacting 
Boson Model (IBM), has been extensively studied by Alhassid and Whelan 
\cite{Alha4, Alha5, Alha6, Whe1, Whe2} and other authors \cite{Liu}, using both classical and quantum measures of chaos. 
In their study, they had found integrability at the three dynamical symmetry limits of 
the symmetry triangle \cite{triangle} of the IBM, namely, U(5), O(6) and SU(3), 
as well as at the O(6)--U(5) side of the triangle, due to the 
O(5) symmetry known  \cite{Talmi} to underlie the O(6)--U(5) line. Away from these integrable 
regions, one expected chaotic behavior. However, the study of the interior of the symmetry 
triangle of the IBM, brought to the surface a region of nearly regular behavior \cite{Alha5,Whe1,Whe2}, 
connecting the U(5) and SU(3) vertices, known as the ``Alhassid--Whelan arc of regularity'' 
(AW arc). 

The increased regularity observed in the region of the Alhassid--Whelan arc, as well as the locus of the arc,
have been studied using several different techniques. 

1) Information entropy of the wavefunctions is a measure that quantifies the eigenstate localization of a particular IBM Hamiltonian in different symmetry bases associated with dynamical symmetries, linking it with the degree of regularity \cite{Cejn1,Cejn2}. It has indeed been found 
\cite{Cejn1,Cejn2} that increased localization in the symmetry bases occurs on the Alhassid--Whelan arc of regularity. 

2) The line corresponding to the degeneracy of the $0_{2}^+$ and $2_{2}^+$ states within the symmetry triangle of the IBM
has been found to closely follow the arc 
of regularity \cite{Jolie,Amon}. Furthermore, more than twelve nuclei exhibiting this behavior were placed 
on the arc, providing an experimental confirmation of its existence \cite{Jolie,Amon}. 
Later on, the approximate degeneracy of the $0_{2}^+$ and $2_{2}^+$ states has been related to the degeneracy of the $\beta$ and $\gamma$ bandheads, its locus has been determined through the intrinsic state formalism, and has been found to be located very close to the Alhassid--Whelan arc \cite{Macek2,Macek80}. 

3) The dynamics of $0^+$ states have also been considered, using the nearest neighbor spacing distribution 
of $0^+$ states, in order to demonstrate the semiregular nature of the arc of regularity, in agreement with 
results obtained using classical measures based on Poincar\'{e} sections \cite{Macek2}. 
In particular, a bunching pattern of $0^+$ states has been found \cite{Macek2} on the arc, similar to the bunching pattern 
seen along the O(6)--U(5) side of the triangle \cite{Heinze}, which is known for its regular dynamics \cite{Talmi}.

4) The line corresponding to the degeneracy of the $2_\beta^+$ and $2_\gamma^+$ states has also been found \cite{Bon1} to closely follow the arc of regularity. In the large boson number limit this degeneracy also guarantees the degeneracies 
predicted by the SU(3) symmetry in the first few bands lying lowest in energy. This can be considered \cite{Bon1} as a sign of an underlying SU(3) quasidynamical symmetry, however its validity is limited to low lying states and to the large boson number 
limit. 

5) A line within the symmetry triangle of the IBM, also lying close to the arc of regularity, has been obtained \cite{Bon2}
using a contraction of the SU(3) algebra to the algebra of the rigid rotator. This finding is related to the ground state 
band alone and has again been obtained in the limit of large boson numbers.

6) Recently, families of high-lying regular rotational bands have been found \cite{Macek3,Macek4} in the IBM framework, occurring even in nuclei far away from the SU(3) dynamical symmetry and leading to increased overall regularity. 
 
The present paper offers a closer view on the quantum chaotic dynamics at the vicinity of the arc of regulatiry,
without attempting  to elucidate the nature of the symmetry underlying the arc.  
Taking advantage of the new code IBAR \cite{IBAR1,IBAR2}, which can handle up to $N_B=1000$ bosons, 
the present study is focused on states with zero angular momentum ($J=0$), but can be easily extended 
to other angular momentum values. The main objectives are described here.

1) To determine any energy dependence of statistical measures of quantum chaos in $0^+$ states and in 
$B(E0)$ transition strengths. This study is for the first time feasible, due to the good statistics allowed
by the large number of bosons used.  

2) To examine the stability of the location of the arc of regularity within the symmetry trianlge of IBM 
with changing boson number. The IBAR code allows for the first time to examine this stability for large boson numbers.
The question of stability is of importance in relation to the empirical evidence \cite{Jolie,Amon} found 
for the arc of regularity, since different nuclei are described by different boson numbers, in the region of deformed 
nuclei already used \cite{Jolie,Amon} the boson number being close to 14. 

In Section II, the IBM Hamiltonian is described. The fluctuation measures, used to 
study the quantum dynamics in the symmetry triangle of the IBM are introduced in Section III, while 
in Section IV  the numerical results are presented. An O(6) line is considered in Section V, while in Section VI the discussion 
of the results is given. 

\section{IBM Hamiltonian and symmetry triangle} %2 

\begin{figure}[htbp]
\centering
\includegraphics[height=60mm]{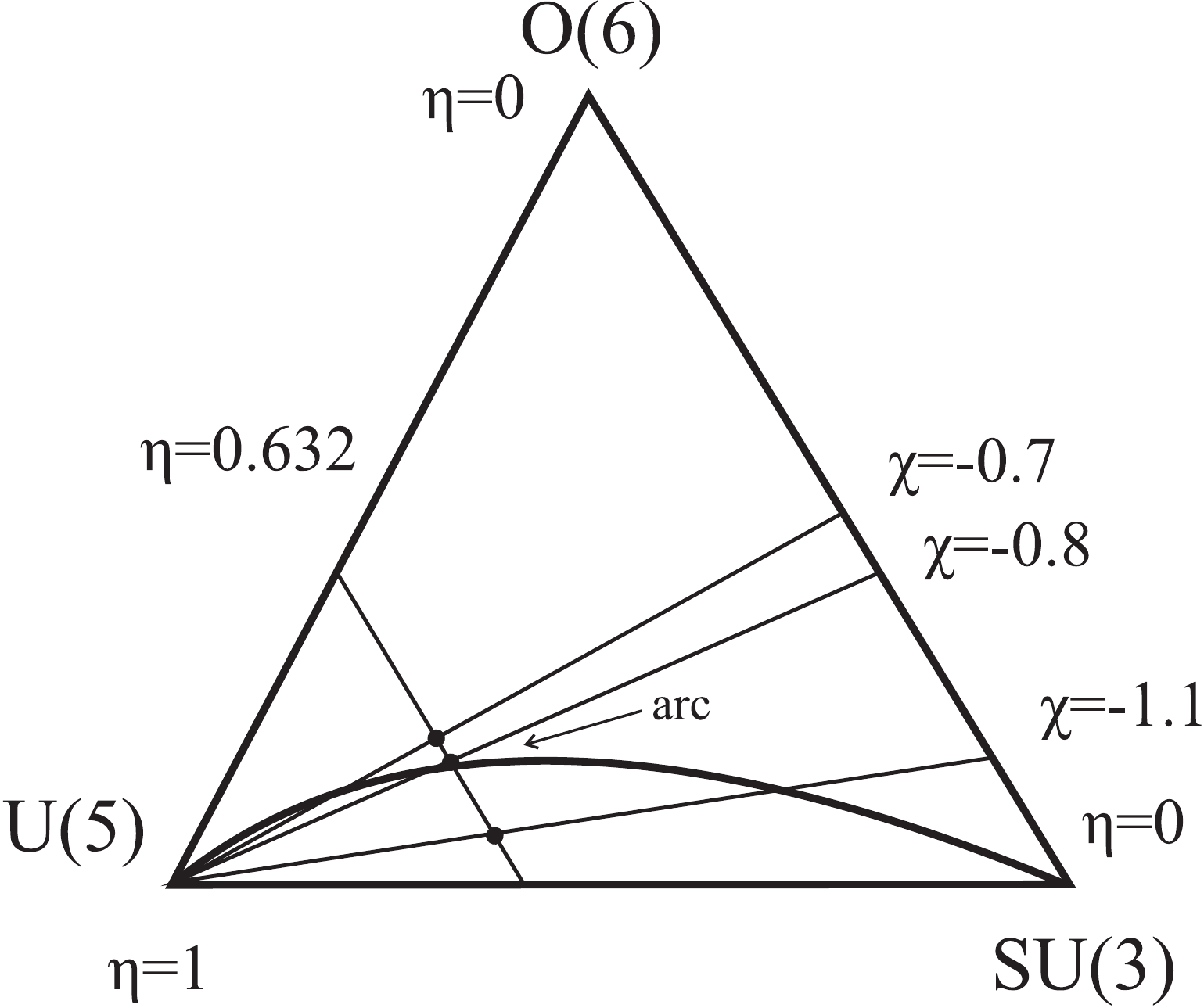}
\caption[]{The IBM symmetry triangle and the points in the vicinity of the arc, where calculations 
were performed.}\label{triangle}
\end{figure}

In the context of the IBM, low lying states in nuclei can be described in terms of a monopole 
boson, $s$, with angular momentum 0 and a quadrupole boson, $d$, with angular momentum 2. The 
36 bilinear combinations ($s^{\dag}s, \; s^{\dag}\tilde{d}_{\mu}, \; d^{\dag}_{\mu}s, 
\; d^{\dag}_{\mu}\tilde{d}_{\nu}$) form a U(6) spectrum generating algebra. The three dynamical 
symmetries of the model, U(5), SU(3) and O(6), which correspond to vibrational, 
rotational, and $\gamma$--unstable nuclei, respectively, are placed at the vertices of the 
symmetry triangle, shown in Figure \ref{triangle}, which is the parameter space of the model. 

In what follows we use the IBM Hamiltonian \cite{Alha5, Whe1, Whe2}, 
\begin{equation}\label{IBMALH}
H\left(\eta,\chi \right)=c \left[ \eta \hat{n}_d +\frac{\eta-1}{N_B}\hat{Q}_{\chi} \cdot \hat{Q}_{\chi}\right],
\end{equation}
where $\hat{n}_d=d^{\dag} \cdot \tilde{d}$ is the d boson number operator, 
$\hat{Q}_{\chi}=(s^{\dag}\tilde{d}+d^{\dag}s)+\chi (d^{\dag} \tilde{d})^{(2)}$ 
is the quadrupole operator, and $N_B$ is the number of valence bosons. 
The parameters $(\eta,\chi)$ are the coordinates of the triangle and serve 
for symmetry breaking. $\eta$ ranges from 0 to 1, and $\chi$ ranges from 
0 to $\frac{-\sqrt{7}}{2} \approx -1.32$. By varying $\eta$ and $\chi$, the three 
dynamical symmetries of the model can be reached. U(5) corresponds to $(\eta,\chi)=(1,0)$, 
O(6) to $(\eta,\chi)=(0,0)$ and SU(3) to $(\eta,\chi)=(0, \frac{-\sqrt{7}}{2})$. 
Numerical calculations of energy levels and $B(E0)$ transition rates 
have been performed using the code IBAR \cite{IBAR1, IBAR2} 
which can handle up to $N_B$=1000 bosons. 

\section{Fluctuation measures} %3 

In this section, the fluctuation measures or statistics, which are applied to the spectrum 
and the transition intensities of $0^+$ states, are introduced. However, before proceeding to 
the quantum chaotic analysis of the spectrum or of the transitions intensities, the eigenvalues 
should be unfolded. The reason is that the Gaussian Orthogonal Ensemble (GOE) requires that the average level spacing, $S$, of a 
spectrum in the limit $N \rightarrow \infty$ 
should be constant, however, the ordered sequence of levels $(E_1, E_2, \dots, E_N)$ produced, for 
example from the IBM Hamiltonian, forms a spectrum in which 
the low energy levels have consistently larger spacings than the high energy ones. In order to be 
consistent with the GOE requirements, one needs to unfold the spectra, that is to say, modify the 
spectrum, so that the average level spacing, $S$, is constant. 
The first step of unfolding is performed by constructing a staircase function of 
the data. A staircase function is the number of levels found below some specific 
energy. Then, a low order polynomial $N(E)$ is fitted to the staircase function \cite{Alha1}. 
The unfolded energies, called normalized energies, are defined as $\epsilon_i=N(E_i)$.
With this mapping, the average level spacing of the spectrum of the normalized energies, 
the unfolded spectrum, becomes constant and specifically, equal to one, $\langle S \rangle=1$. 
The fluctuation measures, are then applied to the unfolded spectrum, which might have constant 
average level spacing, however, the spacings still show strong fluctuations. 

The statistical measures used for the determination of the fluctuation properties of the 
unfolded spectrum are the nearest neighbor level spacing distribution $P(S)$ \cite{Brody} and the $\Delta_3$ 
statistics of Dyson and Mehta \cite{Dyson4,Mehta}. For the quantum statistical analysis of the 
transition intensities, the distribution $P(y)$ \cite{Levine,Feingold}, a $\chi^2$ distribution, 
with $\nu$ degrees of freedom is applied to the $B(E0)$ transition intensities.

$P(S)$ is defined as the probability that two adjacent energies 
differ by an amount of $S$. First, the normalised spacings, 
$S_i=\epsilon_{i+1}-\epsilon_i$, are calculated and placed into bins, so a histogram of 
normalised spacings is produced. Then, the Brody distribution \cite{Brody} is fitted to the histogram 
\begin{equation}\label{Brody} 
P_{\omega}(S)=A \, \alpha (1+\omega)S^{\omega} \exp(-\alpha S^{1+\omega}),
\end{equation} 
where $A$ is a scaling factor and $\alpha=\Gamma[(2+\omega)/(1+\omega)]^{1+\omega}$. 
The value of $\omega$ is found by fitting Eq. (\ref{Brody}) to the data by least squares. The 
Brody distribution takes the form of Poisson statistics when $\omega=0$, which 
characterise a regular system,  and of the Wigner distribution when $\omega=1$, which 
corresponds to a chaotic system. For intermediate cases, larger $\omega$ values imply 
more chaos. 

Spectral rigidity, $\Delta_3(L)$, is a measure of the deviation of the staircase function 
from a straight line, 
used to measure long range correlations. It was introduced by Dyson and Mehta \cite{Dyson4,Mehta}, who 
defined the function 
\begin{equation}
\Delta_3(a, L)=\frac{1}{L}min_{A, B} \int^{a+L}_a [N(E)-(AE+B)]^2dE, 
\end{equation}
where the constants $A$ and $B$ will give the best local fit to $N(E)$ in the interval 
$a \leq E < a+L$ and $L$ is the energy length of the interval. 
For a random Poisson spectrum (regular case), $\Delta_3$ takes the form 
\begin{equation}\label{DPoisson}
\Delta_{3}^{P}(L)=\frac{L}{15}.
\end{equation}
For the GOE (chaotic) case there is an approximate expression, for large $L$,
\begin{equation}\label{DGOE}
\Delta_{3}^{GOE}(L)=\frac{1}{\pi^2}(logL-0.0687).
\end{equation}
The exact expression, good for all $L$, is also known \cite{Bohigas1}. In fact, in order to calculate 
the $\Delta_3$ statistics, in terms of the normalised 
energies $\epsilon_1, \epsilon_2, \dots, \epsilon_n$, the function of $\Delta_3(a, L)$ is used, 
as given in Eq. (22) in \cite{Alha1}, 
\begin{eqnarray}
\Delta_3(\alpha,L) &=& \frac{n^2}{16} -\frac{1}{L^2} \left( \sum_{i=1}^n \tilde{\epsilon}_i \right)^2 + \frac{3n}{2L^2} \left( \sum_{i=1}^n \tilde{\epsilon}_i^2 \right)\nonumber \\
             -\frac{3}{L^4} \left( \sum_{i=1}^n \tilde{\epsilon}_i^2 \right)^2   & + &\frac{1}{L} \left(\sum_{i=1}^n (n-2i+1) \tilde{\epsilon}_i \right),
\end{eqnarray}
where $\tilde{\epsilon}_i=\epsilon_i-(\alpha+L/2)$ is the measure of the normalised 
energies with respect to the center of the energy interval ($\alpha, \alpha+L$).
$\Delta_3(L)$ is calculated in energy intervals of length 
$L$, which span the whole normalised spectrum, 
once for intervals starting at $\alpha=0$ and once for 
intervals starting at $\alpha=L/2$. Then, $\Delta_3(L)$ 
is found as the average over all $\Delta_3(\alpha,L)$. 
The form of the function that should be fitted on the data is \cite{Brody}
\begin{equation}\label{D3}
\Delta_{3}^{q}(L)=\Delta_{3}^{GOE}(qL)+\Delta_{3}^{P}((1-q)L).
\end{equation}
The value of $q$ is again found from the fitting of Eq. (\ref{D3}) to the data points. 
For $q=0$, the regular case is reached, 
$\Delta_{3}^{P}(L)$, while for $q=1$, the chaotic limit emerges, $\Delta_{3}^{GOE}(L)$. 
For intermediate values of $q$, the behavior of the system is closer 
to chaos as $q$ is closer to 1. 

The last distribution, $P(y)$, where $y$ is the relevant transition intensity, 
e.g. $B(E0)$, is constructed in such a way that $P(y)dy$ 
is the probability of finding an intensity in the interval $dy$ around $y$.
After proper normalization of the transition strengths $y$ (see \cite{Whe1, Whe2, Alha1}), 
their logarithms are assigned to bins and a histogram of normalized transition strengths 
is produced. Then, the interpolating function \cite{Levine,Feingold}
\begin{equation}\label{Porter}
P_{\nu}(y)=A \, \left( \frac{\nu}{2<y>} \right)^{\nu/2} \frac{y^{\frac{\nu}{2}-1}exp(-\nu y/2 <y>)}{\Gamma \left( \frac{\nu}{2}\right) },
\end{equation} 
where $A$ is a factor added for scaling reasons, is fitted to the histogram, through a 
least squares fitting, in order to find the best value of $\nu$. However, since $y$ is 
the logarithm of the normalized transition strengths, one should change the variable $y$ 
of the intepolating function of Eq. (\ref{Porter}) to $z=\log_{10}(y)$ and use for the fitting 
the form of the interpolating function after the change of variable. When $\nu=1$, the 
interpolating function reduces to the Porter--Thomas distribution \cite{stat5}, which 
is the GOE case (chaotic),
\begin{equation}
P_{GOE} (y)=\frac{1}{\sqrt{2\pi <y>}}\frac{1}{\sqrt{y}}exp(-y/2<y>).
\end{equation}
For small values of $\nu$ regularity is expected. There is no formal expression for the 
regular case.

\section{Numerical results} %4 

\subsection{Quantum chaotic dynamics of $0^+$ states}

Measures of chaotic dynamics were calculated at four different points in the IBM 
symmetry triangle, as is shown in Fig. \ref{triangle}, 
namely on the SU(3) vertex, a point on the arc of regularity having parameters 
$(\eta, \chi)=(0.632, -0.803)$ and at two points with the same $\eta$ lying off the arc 
at $(\eta, \chi) = (0.632, -0.7)$ and $(\eta, \chi)=(0.632, -1.1)$. The value $\eta=0.632$ corresponds 
to $\zeta=0.7$ in a different parametrization \cite{Bon2}, and was chosen in order to be in accordance with the value 
used in Ref. \cite{Bon1}. The region of coexistence begins at $\eta=0.8$~. The points on the arc 
are given by the expression $\chi(\eta)=\frac{\sqrt{7}-1}{2}\eta-\frac{\sqrt{7}}{2}$, 
which was found by calculating the values of $\chi$ versus $\eta$, where $\sigma$ 
(a measure of classical chaos) is minimized \cite{Cejn2}. 

\begin{figure*}[htb]
\centering
\includegraphics[height=110mm]{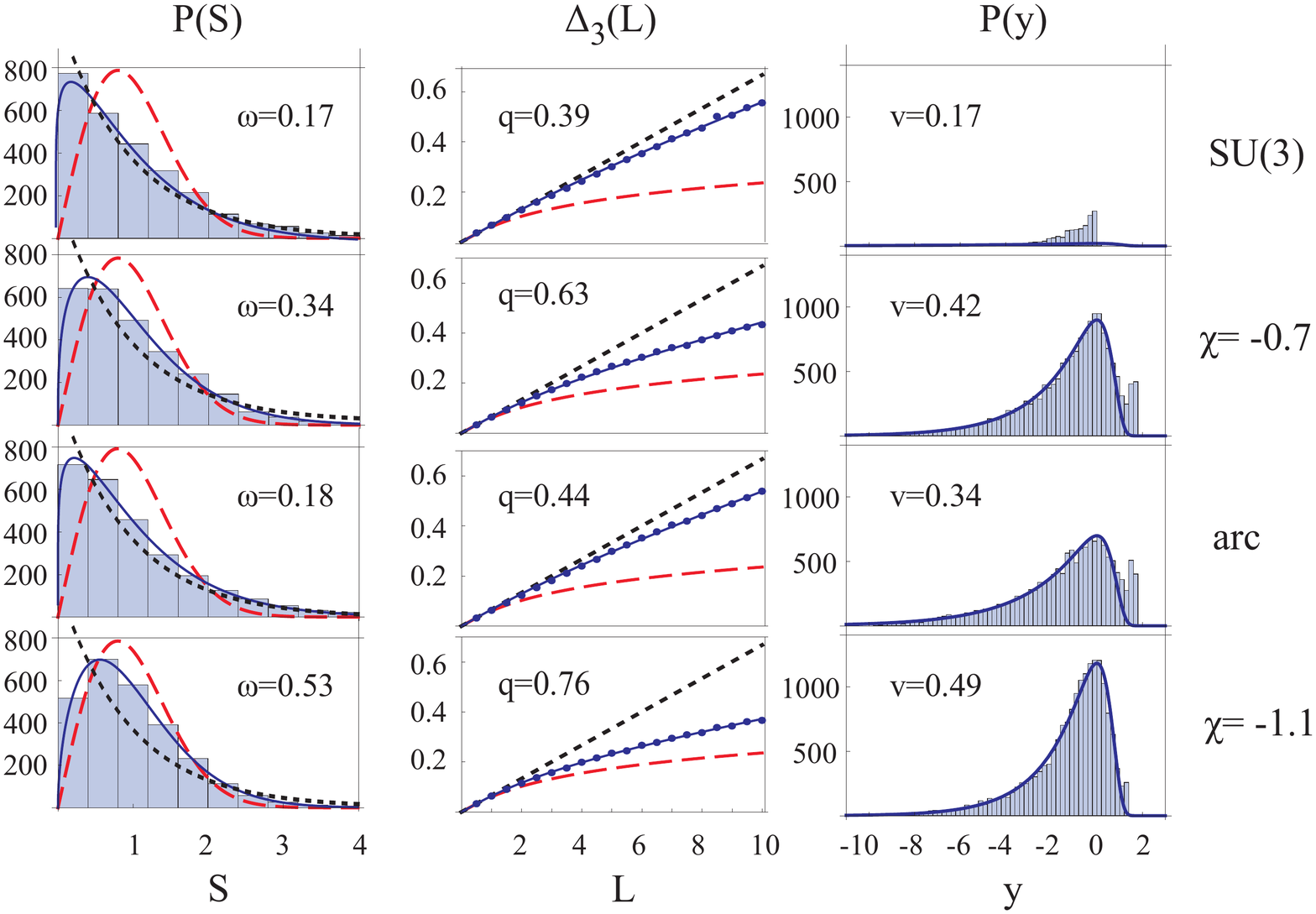}
\caption[]{(color online) Results for the four points shown in Fig. \ref{triangle}, for $N_B=175$ 
and $J=0$ states, using the fluctuation 
measures $P(S)$ of Eq. (\ref{Brody}) \cite{Brody} and $\Delta_3(L)$ of Eq. (\ref{D3}) \cite{Dyson4,Mehta},
and for $N_B=50$  and $J=0$ states, for the fluctuation measure $P(y)$ of Eq. (\ref{Porter}) \cite{Levine,Feingold}. 
The black dotted line shows the regular limit, the red dashed line describes the chaotic limit, 
while the blue solid line is the fit 
to the distribution. It is evident from the values of $\omega$, $q$ and $\nu$, that the SU(3) vertex and 
the point on the arc are more regular compared to the other two points above and below the 
arc.}\label{four}
\end{figure*}

In order to have good statistics, $N_B=175$ was used for the fluctuation measures 
$P(S)$ and $\Delta_3(L)$, producing 2640 $0^+$ states. For the $P(y)$ fluctuation measure, 
$N_B=50$ was used, producing 54,756 possible transition strengths, 
$B(E0)$s, between the 234 $0^+$ states. The reason for selecting $N_B=50$ 
instead of $N_B=175$, was that in the latter case more than 6,000,000 possible transition intensities 
are produced  
between the 2640 $0^+$ states, which renders the calculation impossible to run in terms 
of time, while for $N_B=50$, the run time and the statistics are 
more than satisfactory. The allowed number of $B(E0)$s differs from point to point 
in the symmetry triangle of the IBM and this number gets larger, as the dynamics of the point 
get more chaotic. Figure \ref{four} shows the results for the three statistical measures 
$P(S)$, $\Delta_3(L)$ and $P(y)$ applied on the three points illustrated in 
Fig. \ref{triangle}, accompanied by the results on the SU(3) vertex, a point with regular 
dynamics.

All three different measures of fluctuations show 
consistent results. The SU(3) vertex is the more regular point, followed by the point 
on the arc of regularity, which appears to also have regular behavior. Then come the 
points labelled by $\chi=-0.7$ and $\chi=-1.1$ which are indeed chaotic, with the $\chi=-0.7$ point 
being less chaotic than the $\chi=-1.1$ point. The fitting of the interpolating function 
of the last statistical measure, $P(y)$, on the SU(3) vertex is rather poor, which reflects the fact 
that the expression of Eq. (\ref{Porter}) does not reduce to the regular case for any value 
of $\nu$. 

\begin{table}[htb]
\caption{Spectra of $0^+$ states obtained from the Hamiltonian of Eq. (\ref{IBMALH}) for $N_B=175$ and 
for the ($\eta$, $\chi$) parameter sets $(0, \frac{-\sqrt{7}}{2})$ [SU(3)], (0.632, $-0.803$) [arc],
(0.632, $-0.7$) and (0.632, $-1.1$). In all cases the first $0^+$ state is set at zero 
and all other energies are normalized to the energy of the second $0^+$ state, i.e. to the energy 
of the first excited $0^+$ state. For each of the 8 intervals of 330 states each, into which the spectrum is divided,
the energy of the highest state in the interval is shown. }\label{T0}
\bigskip
\begin{center}
\begin{tabular}{r r r r r r r r r }
state &         &SU(3)  & & arc & & $\chi=-0.7$ & & $\chi=-1.1$  \\  \hline

   1 & $\qquad$ &     0 & $\quad$ &     0 & $\quad$ &     0 & $\quad$ &     0           \\
   2 & $\qquad$ &     1 & $\quad$ &     1 & $\quad$ &     1 & $\quad$ &     1           \\
 330 & $\qquad$ & 29.19 & $\quad$ & 26.84 & $\quad$ & 25.39 & $\quad$ & 28.64           \\
 660 & $\qquad$ & 38.14 & $\quad$ & 37.37 & $\quad$ & 36.52 & $\quad$ & 38.38           \\
 990 & $\qquad$ & 43.30 & $\quad$ & 47.02 & $\quad$ & 46.82 & $\quad$ & 47.32           \\
1320 & $\qquad$ & 46.49 & $\quad$ & 56.16 & $\quad$ & 56.57 & $\quad$ & 55.74           \\
1650 & $\qquad$ & 49.58 & $\quad$ & 64.90 & $\quad$ & 65.94 & $\quad$ & 63.78           \\
1980 & $\qquad$ & 52.74 & $\quad$ & 73.42 & $\quad$ & 75.00 & $\quad$ & 71.54           \\
2310 & $\qquad$ & 55.87 & $\quad$ & 81.66 & $\quad$ & 83.79 & $\quad$ & 79.16           \\
2640 & $\qquad$ & 59.00 & $\quad$ & 92.74 & $\quad$ & 94.28 & $\quad$ & 91.17           \\ 
  
\end{tabular}
\end{center}
\end{table}

\begin{figure*}[htb]
\centering
\includegraphics[height=185mm]{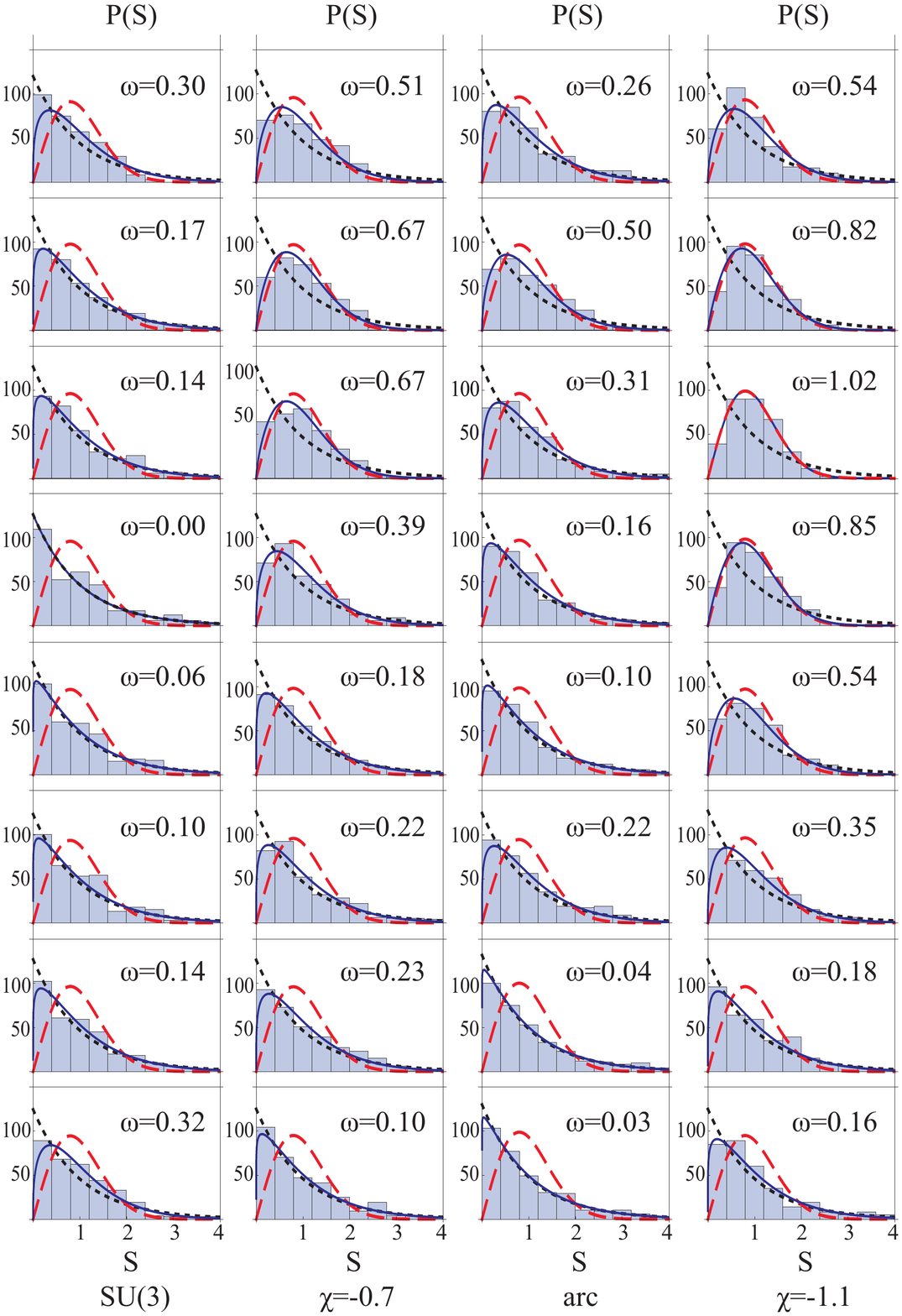}
\caption[]{(color online) Results obtained for the nearest spacing distribution P(S) of Eq. (\ref{Brody}) \cite{Brody}
 for each of the cases shown in Fig. \ref{triangle}, for $N_B=175$ and,
 when the set of $J=0$ states is divided into 8 sets 
of 330 states each. The states 1-330 (labelled as interval 1 in Table \ref{T1}) appear on 
the top, the states 2311-2640 (labelled as interval 8 in Table \ref{T1}) appear at the 
bottom. The black dotted line shows the regular limit, the red dashed line describes the chaotic limit, 
while the blue solid line is the fit to the distribution.}\label{PSparts}
\end{figure*}

\begin{figure*}[htb]
\centering
\includegraphics[height=185mm]{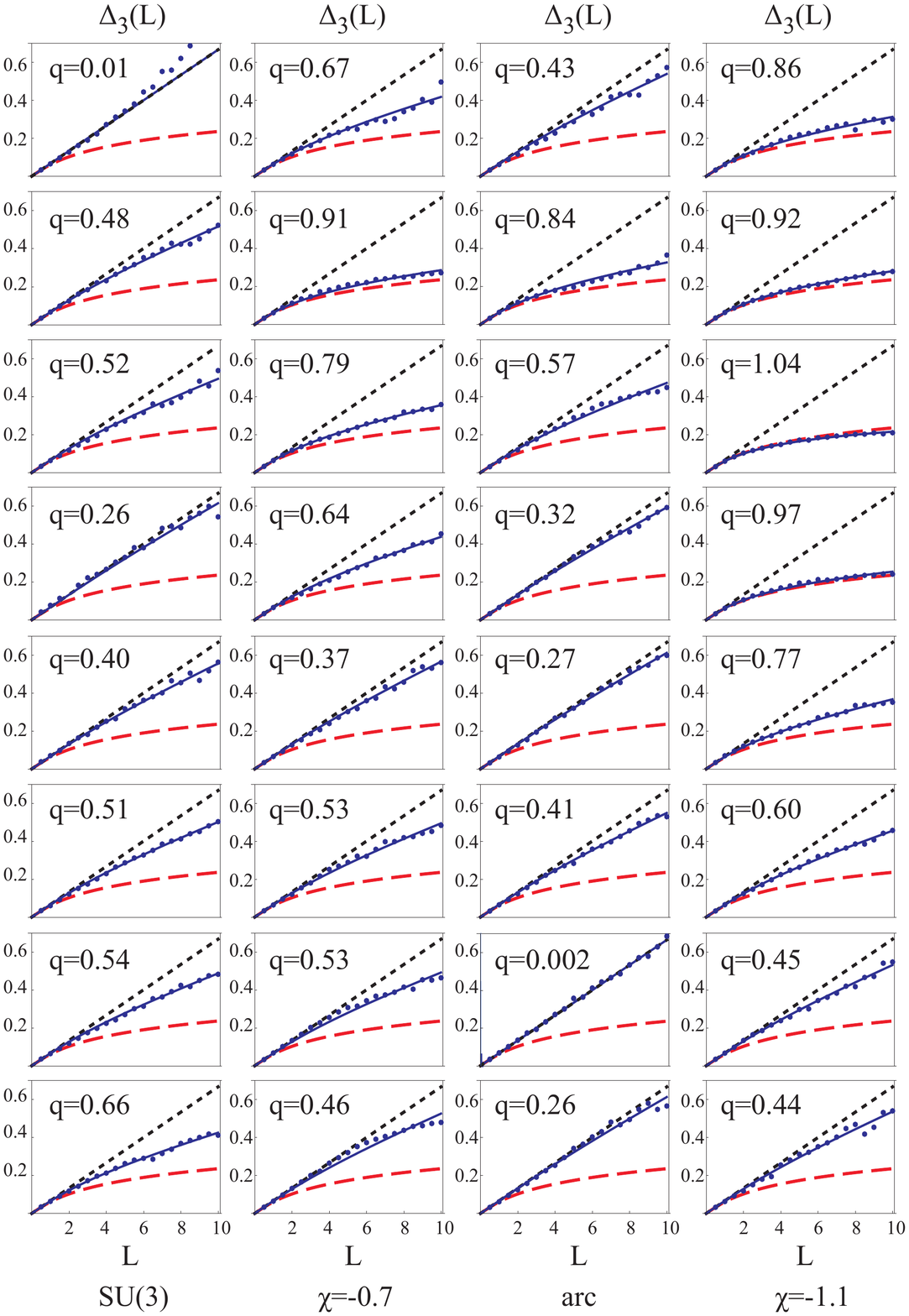}
\caption[]{(color online) Results obtained for the spectral rigidity measure $\Delta_3^q(L)$
of Eq. (\ref{D3}) \cite{Dyson4,Mehta}, 
for each of the cases shown in Fig. \ref{triangle}, 
for $N_B=175$ and, when the set of $J=0$ states is divided into 8 sets 
of 330 states each. The states 1-330 (labelled as interval 1 in Table \ref{T2}) appear on 
the top, the states 2311-2640 (labelled as interval 8 in Table \ref{T2}) appear at the 
bottom. The black dotted line shows the regular limit, the red dashed line describes the chaotic limit, 
while the blue solid line is the fit to the distribution.}\label{D3parts}
\end{figure*} 

\begin{table}[htb]
\caption{Numerical values of the parameter $\omega$ of the Brody distribution $P_\omega(S)$ of Eq. (\ref{Brody}), 
for the 8 parts of each spectrum for $N_B=175$. The states 1-330 are labelled 
as interval 1, the states 2311-2640 are labelled as interval 8.
The value of $\omega$ obtained from fitting the whole spectrum, coming from Fig. 2, 
is labelled by ``total''.}\label{T1}
\bigskip
\begin{center}
\begin{tabular}{c c c c c c c c c}
interval & &SU(3) & & arc & & $\chi=-0.7$ & & $\chi=-1.1$  \\ 
P(S)&$\qquad$& $\omega$&$\quad$ & $\omega$&$\quad$ & $\omega$ &$\quad$ & $\omega$  \\ \hline
1 & $\qquad$ & 0.30 & $\quad$ & 0.26 & $\quad$ & 0.51 & $\quad$ & 0.54           \\
2 & $\qquad$ & 0.17 & $\quad$ & 0.50 & $\quad$ & 0.67 & $\quad$ & 0.82           \\
3 & $\qquad$ & 0.14 & $\quad$ & 0.31 & $\quad$ & 0.67 & $\quad$ & 1.02           \\
4 & $\qquad$ & 0.00 & $\quad$ & 0.16 & $\quad$ & 0.39 & $\quad$ & 0.85           \\
5 & $\qquad$ & 0.06 & $\quad$ & 0.10 & $\quad$ & 0.18 & $\quad$ & 0.54           \\
6 & $\qquad$ & 0.10 & $\quad$ & 0.22 & $\quad$ & 0.22 & $\quad$ & 0.35           \\
7 & $\qquad$ & 0.14 & $\quad$ & 0.04 & $\quad$ & 0.23 & $\quad$ & 0.18           \\
8 & $\qquad$ & 0.32 & $\quad$ & 0.03 & $\quad$ & 0.10 & $\quad$ & 0.16           \\ 
  & $\qquad$ &      & $\quad$ &      & $\quad$ &      & $\quad$ &                \\
total&$\qquad$&0.17 & $\quad$ & 0.18 & $\quad$ & 0.34 & $\quad$ & 0.53           \\ 

\end{tabular}
\end{center}
\end{table}

\begin{table}[htb]
\caption{Numerical values of the parameter $q$  of the $\Delta^q_3(L)$ distribution of Eq. (\ref{D3}) \cite{Dyson4,Mehta}, 
for the 8 parts of each spectrum for $N_B=175$. The states 1-330 
are labelled as interval 1, the states 2311-2640 are labelled as interval 8.
The value of $q$ obtained from fitting the whole spectrum, coming from Fig. 2, 
is labelled by ``total''.}\label{T2}
\bigskip
\begin{center}
\begin{tabular}{ c c c c l c c c c} 
interval & $\qquad$ & SU(3) & $\quad$ & arc & $\quad$ & $\chi=-0.7$ & $\quad$ & $\chi=-1.1$  \\ 
$\Delta_3$(L)& $\qquad$ & $q$& $\quad$ & $q$& $\quad$ & $q$ & $\quad$ & $q$  \\ \hline
1 & $\qquad$ & 0.01 & $\quad$ & 0.43 & $\quad$ & 0.67 & $\quad$ & 0.86          \\
2 & $\qquad$ & 0.48 & $\quad$ & 0.84 & $\quad$ & 0.91 & $\quad$ & 0.92           \\
3 & $\qquad$ & 0.52 & $\quad$ & 0.57 & $\quad$ & 0.79 & $\quad$ & 1.04           \\
4 & $\qquad$ & 0.26 & $\quad$ & 0.32 & $\quad$ & 0.64 & $\quad$ & 0.97           \\
5 & $\qquad$ & 0.40 & $\quad$ & 0.27 & $\quad$ & 0.37 & $\quad$ & 0.77           \\
6 & $\qquad$ & 0.51 & $\quad$ & 0.41 & $\quad$ & 0.53 & $\quad$ & 0.60           \\
7 & $\qquad$ & 0.54 & $\quad$ & 0.002& $\quad$ & 0.53 & $\quad$ & 0.45           \\
8 & $\qquad$ & 0.66 & $\quad$ & 0.26 & $\quad$ & 0.46 & $\quad$ & 0.44           \\ 
  & $\qquad$ &      & $\quad$ &      & $\quad$ &      & $\quad$ &                \\
total & $\qquad$ & 0.39 & $\quad$ & 0.44 & $\quad$ & 0.63 & $\quad$ & 0.76           \\  
\end{tabular}
\end{center}
\end{table}

\subsection{Chaos in $0^+$ states as a function of energy}

The study of chaos as a function of energy, using the quantum spectrum, is for the first time 
possible, due to the large number of bosons used, which allows for good statistics. The 
spectrum of the produced states is divided into equal parts and each part is studied using the 
three statistical measures. For $N_B=175$, 2640 $0^+$ states are 
produced. The spectrum is divided into 8 parts of 330 states and each part is studied separately 
for its chaotic dynamics. The highest energy in each interval is shown in Table \ref{T0}. In all cases 
the first $0^+$ state is set at zero energy, while all other energies are normalized to the energy of the second $0^+$ state,
i.e. to the first excited $0^+$ state, thus rendering the parameter $c$ appearing in Eq. (\ref{IBMALH}) irrelevant.  

The spectra shown in Table \ref{T0} are the original ones, before any unfolding is applied to them. 
The division of the spectrum in 
parts of equal number of states, is the same as the division of the spectrum in parts 
having equal energy difference, due to the normalization of the spectrum which has 
led to neighboring energies differing on the average by 1. Figures \ref{PSparts} and \ref{D3parts} illustrate 
the results obtained for the nearest spacing distribution, P(S), and the spectral rigidity 
measure, $\Delta_3$(L), respectively. The first energy interval, i.e. the part with the states 
1-330, appears at the top of the columns, while the last energy interval, the part with the 
states 2311-2640, appears at the bottom. The numerical results of $\omega$ and $q$ are 
displayed in Tables \ref{T1} and \ref{T2}. 

The degree of chaos is not uniform in energy. The low energy part of the spectrum (the first 
energy interval) is always less chaotic than the closest higher parts of the spectrum 
(the second and third energy intervals), where the motion becomes apparently chaotic. However, 
at higher energies chaos decreases significantly, the spectrum becoming almost regular at its highest 
part, even for the most chaotic point, $\chi=-1.1$. This behavior is common 
at all three points located at the vicinity of the arc. However, the SU(3) point, which is 
of course regular, seems not to follow the behavior of the others. Indeed, chaoticity falls 
in the middle part of the spectrum, and rises in the highest part, displaying exactly the  
opposite properties, although on the average the SU(3) point is less chaotic than the others,
as indicated by the lowest values of $\omega$, $q$ associated with it in Fig.~2. Another observation is that the point on the arc is always less chaotic, 
in all energy intervals, compared to the points above and below the arc and that chaotic 
behavior is confined to fewer intervals at the arc or the SU(3) vertex. 
 
These results are in accordance with several previous studies of both classical chaos and  quantum  chaos; for the latter non-statistical measures of chaos have been used so far. A few examples are listed here.
 
1) The energy dependence of regularity in the classical IBM has been studied for $\eta=0.5$ in Ref. \cite{Macek2}. 
Increased regularity has been found at low energies and again at high energies, while reduced regularity 
has been observed in between. 

2) The energy dependence of regularity for a quantum IBM Hamiltonian has been studied, again for $\eta=0.5$, in Ref. \cite{Macek80}. Increased regularity has been observed in the AW arc region, both at low and at high energies. 
In this quantum study a visual method originally proposed by Peres \cite{Peres} has been used, enabling a qualitative distinction between regular and chaotic motion. 

3) The energy dependence of regularity in a quantum IBM Hamiltonian has been studied 
in a region corresponding to axially deformed ground states in Ref. \cite{Macek3}. Increased regularity, with strong occurence of SU(3)-like rotational bands, has been found at the AW arc at low energies and again at high energies, with reduced regularity occurring in between. Again the visual method of Peres \cite{Peres} has been used. 

\begin{figure}[htb]
\centering
\includegraphics[height=60mm]{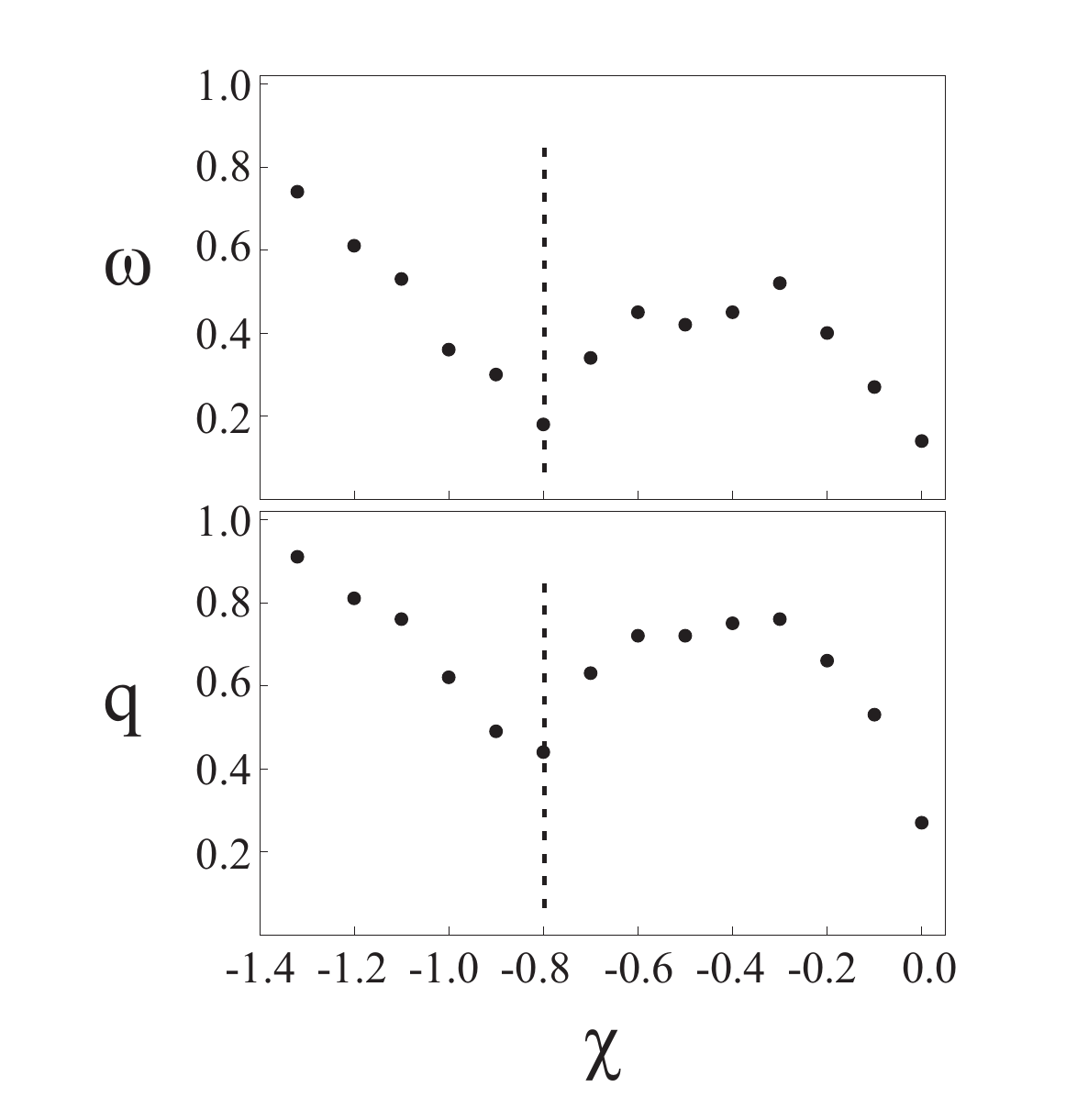}
\caption[]{Results obtained for the quantum statistical parameters $\omega$ and $q$,
for 14 different values of $\chi$ along the $\eta=0.632$ line of the triangle. The 
calculations were performed for $N_B=175$ and $J=0$.}\label{wqvsx}
\end{figure} 

\subsection{Chaos in $0^+$ states as a function of the parameter $\chi$}

The results of the quantum statistical parameters, $\omega$ and $q$, as a function 
of the parameter $\chi$, determined at 14 different points along the $\eta=0.632$ line of the triangle, are seen 
in Figure \ref{wqvsx}. First, being on the SU(3)--U(5) line of the triangle 
($(\eta,\chi)=(0.632, -1.32)$) chaotic behavior is displayed, which keeps 
diminishing as one reaches the arc of regularity 
($(\eta,\chi)=(0.632, -0.803)$), where there is a minimum. Then, as one 
moves to larger values of $\chi$, chaoticity emerges again, but once more gives its 
place to regular behavior as one reaches the O(6)--U(5) line of the triangle 
($(\eta,\chi)=(0.632, 0)$), where the O(5) symmetry causes integrability \cite{Talmi}. 

The results for the $0^+$ states, are in complete agreement with the original work 
of Alhassid and Whelan, who found the existence of the arc of regularity using 
$N_B=25$ and $J\geq 2$ angular momentum. A strong peak on the arc of 
regularity has also been noticed in Ref. \cite{Macek2}, where the authors studied the 
dependence on $\chi$ at $\eta$=0.5 for the ratio of the number of regular 
trajectories to the total number of trajectories.

\begin{figure}[htb]
\centering
\includegraphics[height=70mm]{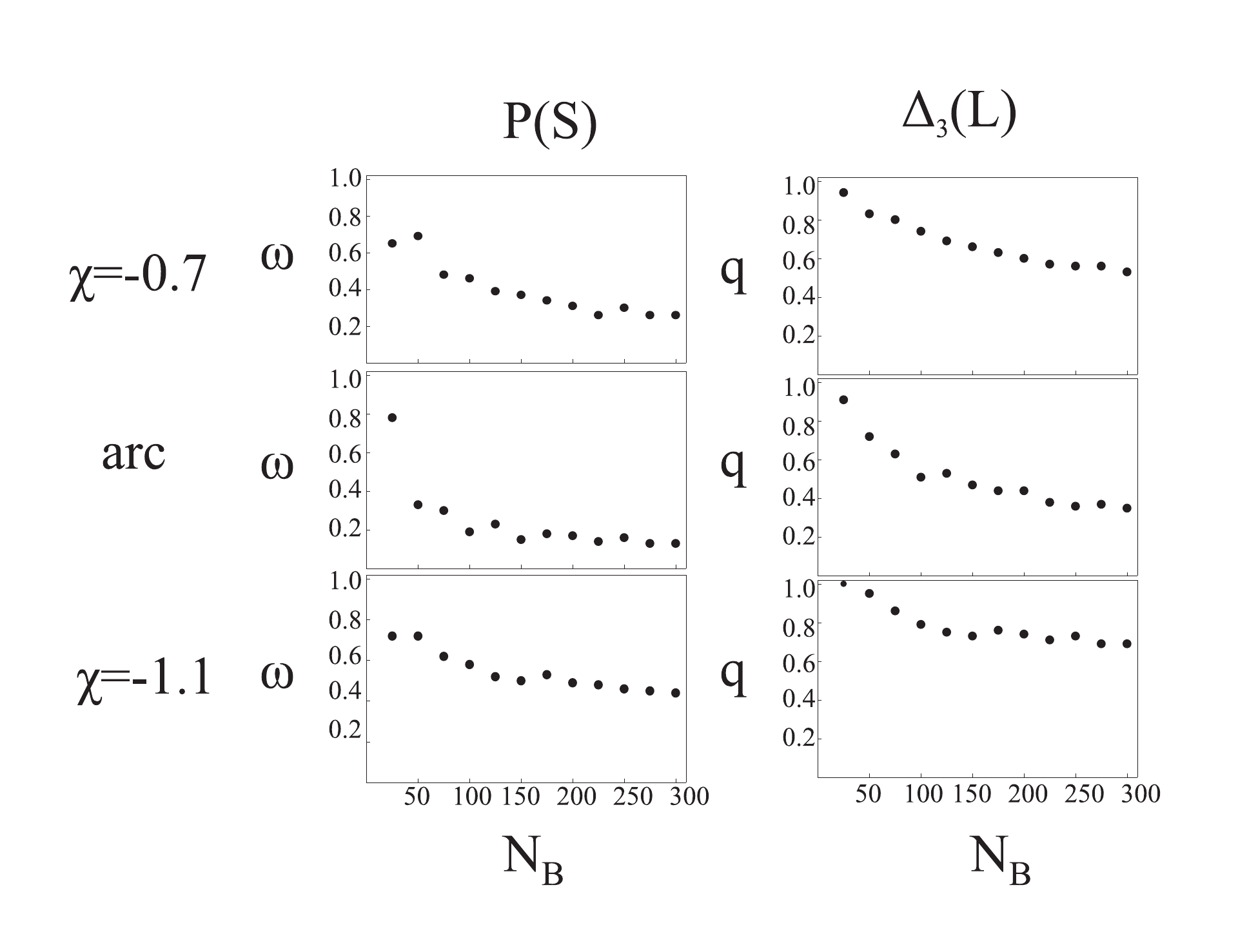} 
\caption[]{Results obtained for the quantum statistical parameters $\omega$ and $q$, 
for various values of $N_B$ at the three points in the vicinity of the arc of regularity.}\label{wqvsN}
\end{figure} 

\subsection{Chaos in $0^+$ states as a function of the number of bosons $N_B$}

The results of the quantum statistical parameters, $\omega$ and $q$, as a function 
of the number of bosons $N_B$ are seen in Figure \ref{wqvsN}. In general, there is a drop 
of the values of the measured quantum 
statistical parameters $\omega$ and $q$ as $N_B$ increases. For 
small number of bosons (till about $N_B=100$) this drop is steep, while for larger 
$N_B$ (larger than $N_B=175$), the drop is very small and the quantum 
statistical parameters seem to have reached steady values. A reason for this steep drop, 
for small values of bosons, can be explained in terms of the total number of states for 
$J=0$, for different number of bosons, seen in Table \ref{T5}, which affects the statistical 
results. For example, for $N_B=25$ there are only 65 states, which give barely sufficient 
statistics, for $N_B=50$ there are 234 states which give better statistics, while for $N_B=175$ 
there are 2640 states which give more validity to the statistical analysis of the eigenvalues.
As the size of the system becomes larger, in the limit $N_B \rightarrow \infty$, 
$\omega$ and $q$ seem to approach fixed asymptotic values.

\begin{table}[htb]
\caption{Total number of states for $J=0$, for different number of bosons $N_B$.}\label{T5}
\bigskip
\begin{center}
\begin{tabular}{ c c c c c c c } 
$N_B$ & $\quad$ & Number of states & $\quad$  & $N_B$ & $\quad$ & Number of states  \\ \hline
25  & $\quad$ &  65   & $\quad$ & 175 & $\quad$ & 2640 \\
50  & $\quad$ &  234  & $\quad$ & 200 & $\quad$ & 3434 \\
75  & $\quad$ &  507  & $\quad$ & 225 & $\quad$ & 4332 \\
100 & $\quad$ &  884  & $\quad$ & 255 & $\quad$ & 5334 \\
125 & $\quad$ &  1365 & $\quad$ &275  & $\quad$ & 6440 \\
150 & $\quad$ &  1951 & $\quad$ & 300 & $\quad$ & 7651 \\
\end{tabular}
\end{center}
\end{table}

Despite the drop of the values of the 
parameters, as a function of $N_B$, the point on the arc has the 
smallest values of $\omega$ and $q$, for all $N_B \geq 50$, compared to the other two points, $\chi=-0.7$ and 
$\chi=-1.1$, the point $\chi=-0.7$ being always less chaotic than $\chi=-1.1$ in the same 
$N_B$ region. The situation is different for $N_B=25$, where the value of $\omega$ is 
greater at the position corresponding to the arc of regularity, than at the neighboring 
points ($\chi=-0.7$ and $\chi=-1.1$). This comes as a surprise, since in their work, Alhassid 
and Whelan had used $N_B=25$ and $J=2$ or $J=10$ in order to locate the arc of regularity. 
Probably, the failure to locate the arc, using the fluctuation measure P(S), in our case, has 
to be attributed to the marginal badness of the statistics, since for $N_B=25$ and $J=0$ there 
are 65 states, while for $N_B=25$ and $J=2$ or $J=10$, there are 117 and 211 states respectively. 
However, as the number of bosons increases, the limiting values are quickly reached and the position
of  the arc of regularity becomes stable. One should recall at this point, that the nuclei found to lie 
close to the arc of regularity \cite{Jolie,Amon} have boson numbers close to $N_B=14$, therefore the location 
of the arc for these nuclei might be different from the one corresponding to the $N_B\to \infty$ limit.

\begin{table}[htb]
\caption{Numerical values of the parameter $\nu$ of the distribution $P_\nu(y)$ of Eq. (\ref{Porter}) \cite{Levine,Feingold}, 
for the 8 parts of each spectrum for $N_B=50$. The states 1-30 are labelled 
as interval 1, the states 211-234 are labelled as interval 8.
The value of $\nu$ obtained from fitting the whole spectrum, coming from Fig. 2, 
is labelled by ``total''.}\label{T3}
\bigskip
\begin{center}
\begin{tabular}{ c c c c c c c } 
interval&$\qquad$ & arc & $\qquad$ & $\chi=-0.7$ & $\qquad$ & $\chi=-1.1$  \\ 
P(y)&$\qquad$ & $\nu$& $\qquad$ & $\nu$ & $\qquad$ & $\nu$  \\ \hline
1  & $\qquad$ & 0.40 & $\qquad$ & 0.62 & $\qquad$ & 0.44          \\
2  & $\qquad$ & 0.48 & $\qquad$ & 0.81 & $\qquad$ & 0.84           \\
3  & $\qquad$ & 0.39 & $\qquad$ & 0.60 & $\qquad$ & 0.87           \\
4  & $\qquad$ & 0.32 & $\qquad$ & 0.49 & $\qquad$ & 0.77           \\
5  & $\qquad$ & 0.29 & $\qquad$ & 0.35 & $\qquad$ & 0.54           \\
6  & $\qquad$ & 0.23 & $\qquad$ & 0.26 & $\qquad$ & 0.37           \\
7  & $\qquad$ & 0.21 & $\qquad$ & 0.21 & $\qquad$ & 0.30           \\
8  & $\qquad$ & 0.21 & $\qquad$ & 0.27 & $\qquad$ & 0.22           \\ 
   & $\qquad$ &      & $\qquad$ &      & $\qquad$ &                \\
total&$\qquad$& 0.34 & $\qquad$ & 0.42 & $\qquad$ & 0.49           \\
\end{tabular}
\end{center}
\end{table}
 
\begin{figure*}[htp]
\centering
\includegraphics[height=185mm]{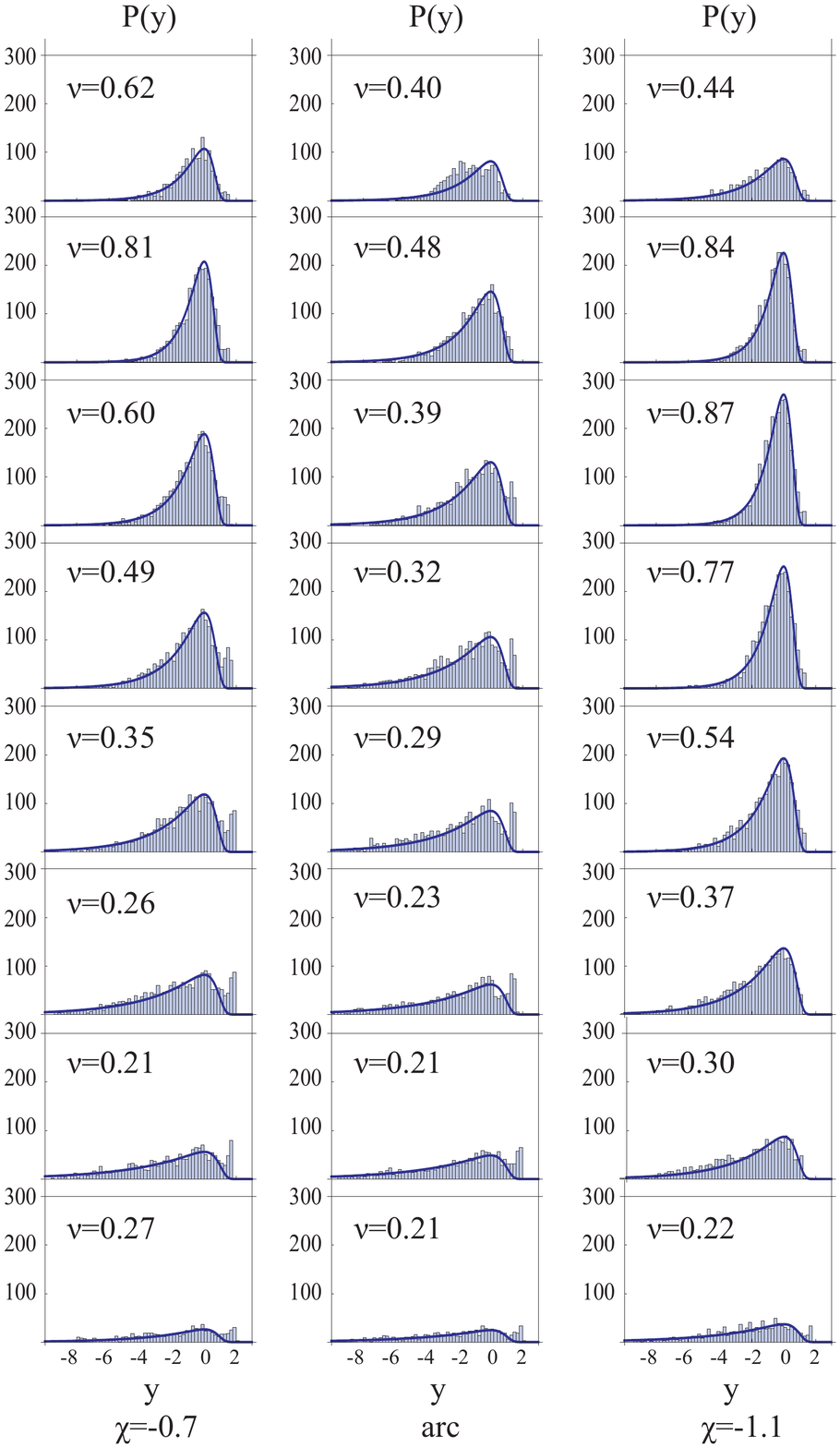} 
\caption[]{(color online) Results obtained for the statistical measure $P(y)$ of Eq. (\ref{Porter}) \cite{Levine,Feingold}, 
for each of the cases shown in Fig. \ref{triangle}, 
for $N_B=50$ and, when the set of 234 $J=0$ states is divided into 7 sets 
of 30 states and one set of 24 states. The states 1-30 
(labelled as interval 1 in Table \ref{T3}) appear on 
the top, the states 211-234 (labelled as interval 8 in Table \ref{T3}) appear at the 
bottom. The blue solid line is the fit to the distribution.}\label{Pypartsparts}
\end{figure*} 

\subsection{Chaos in $B(E0)$ intensities as a function of energy}

In the following subsections, a quantum chaotic study is carried out for the transition 
intensities between the $0^+$ states. The chaotic behavior of $B(E0)$ intensities 
as a function of energy is first presented. The 234 $0^+$ states occuring for $N_B=50$ were divided into 
7 parts of 30 states each and one part of 24 states. For each interval of 30 $0^+$ 
states, the statistical measure $P(y)$ was applied to the $B(E0)$ intensities, 
produced by these 30 $0^+$ states. The results are displayed in Table \ref{T3} and in 
Figure \ref{Pypartsparts}. The first energy interval, i.e. the part with the states
1-30, appears at the top of the columns, while the last energy interval, the part with the
states 211-234, appears at the bottom.  
The SU(3) point is missing from  Table \ref{T3}, for the following reason. As already 
mentioned, for points possessing some dynamical symmetry and thus characterized by regularity, as the SU(3) point, 
 the interpolating function of Eq. (\ref{Porter}) is 
poorly fitted, since the number of $B(E0)$ intensities occurring in this case is much smaller 
than the number of $B(E0)$ intensities obtained at other points, characterized by chaotic 
dynamics. In addition, these relatively few $B(E0)$ intensities are greatly dispersed.  
As a consequence, the study of chaos as a function of energy in the 
SU(3) limit, using $N_B=50$, was statistically impossible. 

Again, the degree of chaos is not uniform in energy. The low energy part of the spectrum (the
first energy interval) is always less chaotic than the next energy interval, where the motion becomes apparently chaotic.
However, at higher energies chaos decreases significantly, the spectrum becoming almost regular at the highest part
of the spectrum, even for the most chaotic point, $\chi = -1.1$. This behavior is common at all
three points located in the vicinity of the arc. Again, the point on the arc displays less 
chaoticity, in all energy intervals,
compared to the points above and below the arc and chaotic behavior is confined to fewer
intervals at the arc. What is fascinating though, is that the study of quantum chaotic dynamics, both on the 
transition intensities and the spectrum of $0^+$ states, show consistency in the degree of 
chaos as a function of energy. In all cases, there is a jump in the degree of 
chaoticity as one passes the low part of the spectum, which is replaced by 
regularity as one gets at the upper part of the spectrum.

\begin{figure}[htb]
\centering
\includegraphics[height=50mm]{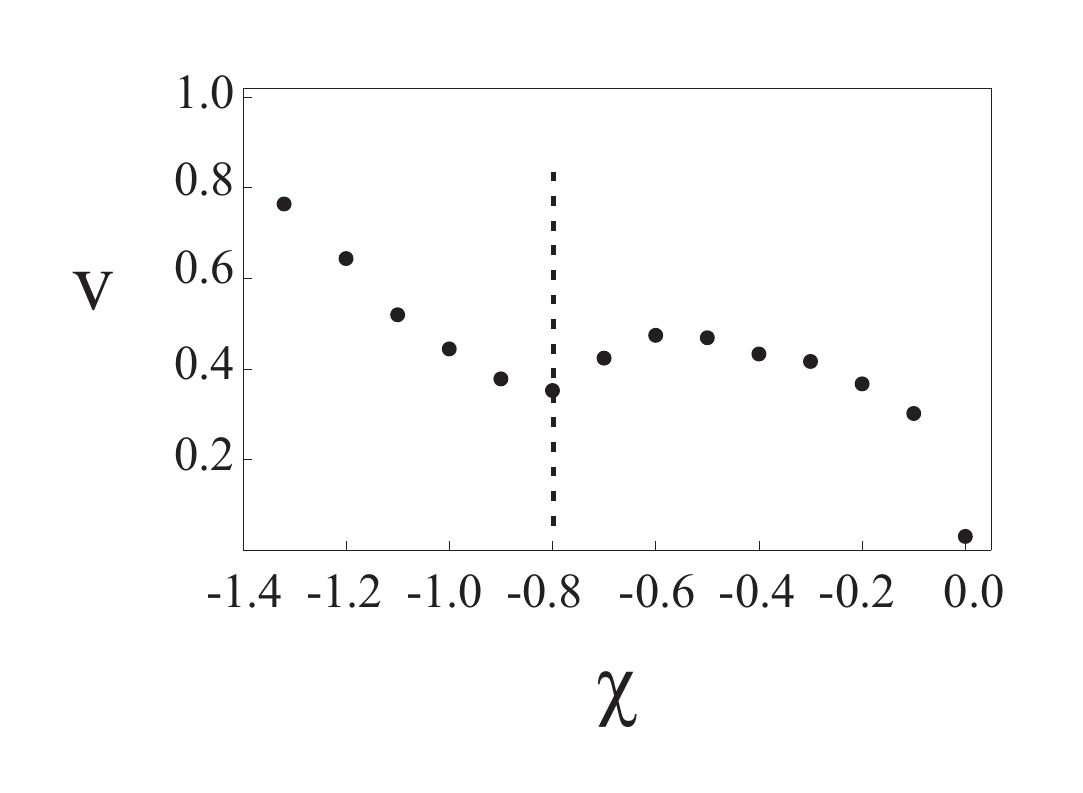} 
\caption[]{Results obtained for the quantum statistical parameter $\nu$,
for 14 different values of $\chi$ along the $\eta=0.632$ line of the triangle. The 
calculations were performed for $N_B=50$ and $J=0$.}\label{Vvschi}
\end{figure}

\subsection{Chaos in $B(E0)$ intensities as a function of the parameter $\chi$}

The results of the quantum statistical parameter $\nu$, as a function 
of the parameter $\chi$, along the $\eta=0.632$ line of the triangle for $N_B=175$ are seen 
in Figure \ref{Vvschi}. For these calculations $N_B=50$ was used, which 
produced 54,756 possible $B(E0)$s between the 234 $0^+$ states. 

The results show the same general behavior, as those of the quantum statistical 
parameters $\omega$ and $q$. There is a characteristic minimum of the curve, at 
$\chi=-0.8$, revealing the minimum chaoticity of the arc. Chaotic behavior 
is encountered at the regions below and above the arc of regularity, while at the 
O(6)--U(5) line of the triangle ($(\eta,\chi)=(0.632, 0)$), regularity emerges 
again. The study of chaoticity of the $B(E0)$ intensities distribution gives results 
which are in complete agreement with the study of chaoticity of the $B(E2)$ intensities 
distribution of the original work of Alhassid and Whelan.

\begin{table}[htb]
\caption{Numerical values for $N_B=175$ of the parameter $\omega$ of the distribution $P_\omega(S)$ of Eq. (\ref{Brody}) \cite{Brody}, 
for the 7 points around the intersection (labelled as ``on arc, on O(6)'') of the O(6) PDS line with the arc of regularity.}\label{T4}
\bigskip
\begin{center}
\begin{tabular}{ l c c c c c c } \hline
 &  & $\chi$ & & $\eta$ & & $\omega$ \\ \hline                 
below arc, on O(6)         & $\qquad$ & -1.10 & $\qquad$ & 0.60 & $\qquad$ & 0.49     \\
on arc, on O(6)            & $\qquad$ & -0.88 & $\qquad$ & 0.54 & $\qquad$ & 0.16     \\
on arc, right of O(6)      & $\qquad$ & -0.98 & $\qquad$ & 0.41 & $\qquad$ & 0.23     \\
on arc, left of O(6)       & $\qquad$ & -0.80 & $\qquad$ & 0.63 & $\qquad$ & 0.18     \\
above arc, on O(6)         & $\qquad$ & -0.50 & $\qquad$ & 0.38 & $\qquad$ & 0.63     \\
above arc, right of O(6)   & $\qquad$ & -0.50 & $\qquad$ & 0.17 & $\qquad$ & 0.55     \\
above arc, left of O(6)    & $\qquad$ & -0.50 & $\qquad$ & 0.57 & $\qquad$ & 0.80     \\ 
\end{tabular}
\end{center}
\end{table}
 
\section{Study of chaos on a line based on O(6) PDS}

In addition to the notion of quasidynamical symmetry (QDS), 
i.e., the approximate persistence of a symmetry, in spite 
of strong symmetry breaking interactions \cite{Rowe1225,Rowe2325,Rowe745,Rowe756,Rowe759}, 
the notion of partial dynamical symmetry (PDS) has been introduced \cite{AL25,Lev77,LVI89,Lev2011},
including three different cases. In type I PDS part of the states possess all the dynamical symmetry,
in type II PDS all the states possess part of the dynamical symmetry, while in type III part of the states 
possess part of the dynamical symmetry \cite{Lev2011}. The linkage between QDS and PDS has been only recently 
clarified, by proving that coherent mixing of one symmetry (QDS)  can lead to partial conservation of a
different, incompatible symmetry (PDS) \cite{Kremer}.  

In Ref. \cite{Kremer}, by using a measure of $\sigma$ fluctuations in a state $\Psi$, called
$\Delta \sigma_{\Psi}$, a valley of almost vanishing
$\Delta \sigma_{g.s.}$ fluctuations has been found, i.e., a valley where the ground state 
wave functions have a high degree of purity with respect to the $\sigma$ quantum 
number of O(6), providing an example of an O(6) approximate PDS of type III. This line 
begins from the O(6) vertex and reaches the U(5)--SU(3) line of the triangle, 
intersecting, at a certain point, with the arc of regularity. 

As a preliminary test, 
we applied the nearest neighbor spacing distribution, on $0^+$ states, for $N_B=175$
bosons, at various points on and around the O(6) PDS line, in order to see whether 
the O(6) PDS line induces regular dynamics to the $0^+$ states considered here.
The results for the relevant parameter, $\omega$, are presented in Table \ref{T4}. The only regular 
points are those located on the arc of regularity, including the intersection of the 
O(6) PDS line with the arc. Therefore, the O(6) PDS does not seem to induce regularity 
to the $0^+$ states considered here, a fact not unexpected, since the O(6) PDS regards 
the ground state band, while the present study focuses on all $0^+$ states, the two sets 
having only one state in common. 

It is expected that in general PDS can lead to suppression of chaos, however this is expected to depend on the number of states 
exhibiting the PDS. In \cite{WAL71} a model Hamiltonian with an SU(3) PDS has been used, in order to examine whether there is suppression of chaos on the point of the PDS. Classical and quantum measures of chaos have been used for three different 
values of the angular momentum ($J=2, 10, 25$) and $N_B=25$. Minimum chaoticity has been found around 
the point where the SU(3) PDS was supposed to exist, but the minimum diverged slightly 
from this point for the two lowest values of angular momentum, while it was more 
pronounced and closer to this point for the larger value of angular momentum ($J=25$). 
The same behavior has been seen for the average SU(3) entropy of the eigenstates of the model 
Hamiltonian, which also showed a more pronounced and closer to the point of the SU(3) PDS 
minimum for $J=25$. This behavior has been explained in terms of the increase of the number 
of soluble states with increasing $J$. In contrast, in \cite{LW77}, where a large 
number of states had an SU(2) PDS, the classical and quantum measures of chaos showed a 
minimum exactly at the point where the SU(2) PDS was supposed to exist.  

\section{Conclusions} %5

In this paper, we presented the application of statistical measures of chaos on the 
energy spectrum and the transition strengths of excited $0^+$ states, in the 
context of the interacting boson model and more precisely on a line connecting 
a point on the  U(5)--SU(3) leg 
of the symmetry triangle with a point on the U(5)--O(6) leg. While the statistical 
measures of chaos 
reveal regularity on the U(5)--O(6) leg, due to the O(5) symmetry known \cite{Talmi} 
to underlie this leg,  
the point on the U(5)--SU(3) leg seems to be the most chaotic point on this 
connection line. Besides the U(5)--O(6) leg, one more minimum is revealed on the 
transition line, located on the arc of regularity, which is in agreement with 
previous studies performed with smaller number of bosons and levels with angular 
momentum $J>2$. 

Concerning the objectives posed at the end of the introduction, the following 
comments are in place. 

1) The study of chaos as a function of energy gave some intriguing results. This 
study has already been performed using both classical  \cite{Macek2} and 
quantum measures of chaos \cite{Macek80,Macek3}, but never using the fluctuation measures 
of the quantum spectrum or the transition intensities, due to the bad statistics imposed by the small number of 
bosons used. The degree of chaos differs as the energy changes. The results, both for 
the spectrum, as well as the transition intensities, show that the low energy part of 
the spectrum has always more regular dynamics compared to the immediately higher 
energy parts, which are almost completely chaotic. However, as the energy increases, 
chaoticity decreases and regularity prevails at the highest parts of the spectrum. 
This is the general behavior for points studied in the vicinity of the arc. 
However, the point on the arc is always more regular, compared to the other points, 
in all energy intervals, giving a confirmation for its regular character. 

2) Quantum chaos also depends on the number of bosons used. As the number of bosons 
increases, the degree of chaos converges to a steady value. Beyond $N_B=50$, the relative 
chaoticity of different points is the same, for all numbers of bosons used in the 
context of this work. For example, chaoticity on the arc points is less than on the neighboring 
points, regardless of the number of bosons beyond $N_B=50$, giving for the arc a steady position in 
the symmetry triangle as a function of boson number beyond $N_B=50$. Below $N_B=50$, however,
the location of the arc appears to be sensitive in $N_B$, this fact affecting the efforts 
of finding nuclei lying on or near the arc \cite{Jolie,Amon}, since the boson numbers 
corresponding to these nuclei are around $N_B=14$.  

While the fluctuation measures of chaos reveal semiregularity on the arc, they cannot reveal 
the nature of the approximate symmetry underlying the Alhassid-Whelan arc of regularity, 
a question which has been addressed in several recent studies \cite{Macek80,Macek3,Bon1,Bon2},
but still remains open. 
 
Recently, a line based on an O(6) PDS was found in the symmetry triangle, extending from the O(6) vertex to a point on the 
U(5)--SU(3) leg of the triangle, intersecting with the arc of regularity. 
Calculations of chaos around the intersection point show that the O(6) PDS does not 
contribute to the development of regular dynamics for the $0^+$ states, a reasonable result
since the O(6) PDS under consideration 
regards only the ground state band of the nuclear spectrum.

The extension of the present study to states with non-zero angular momentum is desirable. 
In this case, attention should be paid to handling degeneracies in the spectrum \cite{Whe2}.  

Study of the interplay of order and chaos has also been extended to the region of shape-phase 
transitions in the triangle. Shape phase transitions between different dynamical symmetries, 
as well as critical point symmeties appearing at the relevant transition points, have been an 
active field of investigation over the last decade \cite{JPG,RMP}. In the framework of the 
interacting boson model, in particular, a first order shape phase transition
between spherical and deformed shapes is known to exist \cite{Feng}, characterized by a 
phase coexistence region \cite{IZC}. The appearance of degeneracies within the phase coexistence 
region \cite{PRL100,PRC80}, as well as the evolution of order and chaos across the first order 
transition \cite{ML84,LM714,LM2097,ML351}, are emerging fields of investigation. 

\section*{Acknowledgments} %6

The authors are grateful to R. J. Casperson, for the code IBAR,  which made the 
present study possible. This work was supported by US Department of
Energy Grant No. DE-FG02-91ER-40609.

\end{document}